\documentclass[12pt,preprint]{aastex}

\newcommand{\msun}{$M_\odot$}
\newcommand{\msunyr}{$M_\odot$~yr$^{-1}$}

\newcommand{\spitzer}{\emph{Spitzer}}
\newcommand{\deltacep}{$\delta$~Cephei}
\newcommand{\dcepb}{HD~213307}
\newcommand{\kms}{km~s$^{-1}$}

\slugcomment{Published on ApJ, 725, 2392 (2010)}

\shorttitle{\deltacep{} Infrared Nebula}
\shortauthors{Marengo et al.}

\begin{document}

\title{An Infrared Nebula Associated with \deltacep:\\
  Evidence of Mass Loss?} 

\author{M. Marengo\altaffilmark{1}, N. R. Evans\altaffilmark{2},
  P. Barmby\altaffilmark{3}, L. D. Matthews\altaffilmark{2,4},
  G. Bono\altaffilmark{5,6}, D. L. Welch\altaffilmark{7},
  M. Romaniello\altaffilmark{8}, D. Huelsman\altaffilmark{9,2},
  K. Y. L. Su\altaffilmark{10} and G. G. Fazio\altaffilmark{2}}
\altaffiltext{1}{Dept. of Physics and Astronomy, Iowa State
  University, Ames, IA 50011}
\altaffiltext{2}{Harvard-Smithsonian Center for Astrophysics, 60
  Garden St., Cambridge, MA 02138} 
\altaffiltext{3}{Dept. of Physics and Astronomy, University of Western
  Ontario, London, Ontario, N6A 4K7 Canada}
\altaffiltext{4}{MIT Haystack Observatory, Off Route 40, Westford, MA 01886}
\altaffiltext{5}{Dept. of Physics, Universit\`a di Roma Tor Vergata,
  via della Ricerca Scientifica 1, 00133 Roma, Italy} 
\altaffiltext{6}{INAF--Osservatorio Astronomico di Roma, via Frascati
  33,  00040 Monte Porzio Catone, Italy}
\altaffiltext{7}{Dept. of Physics and Astronomy, McMaster University,
  Hamilton, Ontario, L8S 4M1, Canada} 
\altaffiltext{8}{European Southern Observatory,
  Karl-Schwarzschild-Str. 2,  85748 Garching bei Munchen, Germany}
\altaffiltext{9}{University of Cincinnati, Cincinnati, OH 45219}
\altaffiltext{10}{Steward Observatory, University of Arizona, 933 N
  Cherry Avenue, Tucson, AZ 85721}

\begin{abstract}

  We present the discovery of an infrared nebula around the Cepheid
  prototype \deltacep{} and its hot companion \dcepb{}. Large scale
  ($\sim 2.1 \cdot 10^4$~AU) nebulosity is detected at 5.8, 8.0, 24
  and 70~\micron{}. Surrounding the two stars, the 5.8 and
  8.0~\micron{} emission is largely attributable to Polycyclic
  Aromatic Hydrocarbon (PAH) emission swept from the ISM by a wind
  originating from \deltacep{} and/or its companion. Stochastically
  heated small dust grains are the most likely source of the 24 and
  70~\micron{} extended emission. The 70~\micron{} emission, in
  particular, resembles a bow shock aligned in the direction of the
  proper motion of \deltacep{}. This discovery supports the hypothesis
  that \deltacep{} may be currently losing mass, at a rate in the
  range $\approx 5 \cdot 10^{-9}$ to $6 \cdot
  10^{-8}$~M$_\odot$~yr$^{-1}$.

\end{abstract}

\keywords{Cepheids --- infrared: stars --- stars: individual:
  \object[del Cep]{$\delta$~Cep} --- stars: mass-loss}

\section{Introduction}\label{sec-intro}

\deltacep{} is a remarkable star. After John Goodricke recognized its
variability in 1784 \citep{goodricke1786}, \deltacep{} has lent its
name to one of the most important classes of variable stars:
Cepheids. Since the discovery of the period--luminosity (PL) relation
(or ``Leavitt law'', \citealt{leavitt1908}), Cepheids and their
prototype \deltacep{} have assumed a fundamental role as the ``first
rung'' of the extragalactic distance scale. Cepheids are also a
crucial benchmark for evolutionary models of intermediate-mass stars
in the He-burning phase. Despite their importance, however, there are
still a number of outstanding puzzles in the theoretical understanding
of the stars pulsating like \deltacep.

The most important issue is the discrepancy between the theoretical
and dynamical mass of Cepheids. First noted by \citet{christy1968},
\citet{stobie1969} and later by \citet{fricke1972}, this issue has
been partially solved \citep{moskalik1992} with the adoption of modern
sets of radiative opacities \citep{seaton1994, rogers1992}. However,
several recent investigations focused on Galactic and Magellanic
Clouds Cepheids \citep{bono2001, bono2002, beaulieu2001, caputo2005,
  keller2006, evans2008} suggest that such a discrepancy still amounts
to 10--15\%. For example, according to \citet{caputo2005} the
pulsational mass of \deltacep{} is $M_P \simeq 4.5$~M$_\odot$, while
the evolutionary mass of the star, based on the
period-color-luminosity relations derived by the same authors, can be
as high as $M_{EV} \simeq 5.7$~M$_\odot$. One possible solution for
this problem is mass loss over the lifetime of these stars. Even
though commonly used semi-empirical relations \citep{reimers1975,
  dejager1997} do not predict enough mass loss to account for the
Cepheids' missing mass, these same relations are clearly inadequate to
describe the mass loss in red giants (see e.g. \citealt{willson2000})
and in globular cluster stars on the Horizontal Branch
\citep{yong2000, castellani2006}. These relations may similarly
underestimate the mass loss rates in intermediate-mass stars like
Cepheids.

Searches for evidence of current or past mass loss by Cepheids and
their progenitors have been conducted at different ends of the
electromagnetic spectrum. In the ultraviolet, \citet{deasy1988}
observed a large sample of Cepheids with the International Ultraviolet
Explorer (IUE) satellite, finding mass loss rates ranging from
$10^{-10}$ to $10^{-7}$~\msunyr. In the far-IR, \citet{mcalary1986}
found evidence in the IRAS photometry \citep{beichman1988} of very
cool dust ($T_d \la 50$~K) around two classical Cepheids (RS~Pup and
SU~Cas) associated with reflection nebulosity. More recently,
\citet{neilson2009} measured mass loss rates of
$10^{-12}$--$10^{-7}$~\msun{} yr$^{-1}$ in a sample of Large
Magellanic Cloud Cepheids in the mid-IR. In the near-IR, $K$ band
interferometric observations detected circumstellar shells around
nearby classical Cepheids \citep{merand2006, merand2007, kervella2006,
  kervella2008, kervella2009}, including \deltacep. A difference in
the absolute distance of $\delta$ Cephei when moving from V, I to V, K
bands was also found by \citep{natale2008} by performing a detailed
fit between predicted and observed light, radius, and radial velocity
curves.

These detections offer tantalizing evidence in favor of Cepheid mass
loss, but are not conclusive. In the case of the far-IR excess, the
large beam of IRAS ($\sim 5$\arcmin) does not allow separation of
local dust emission from background ``Galactic cirrus''. The near-IR
shells detected with interferometers have radii as small as $\sim 2$
stellar radii \citep{merand2006}, and their origin is not well
understood. To resolve these issues, we have obtained \spitzer{} Space
Telescope \citep{werner2004a, gehrz2007} observations of 29 Galactic
Classical Cepheids with the Infrared Array Camera (IRAC,
\citealt{fazio2004}) and the Multiband Infrared Photometer for
\spitzer{} (MIPS, \citealt{rieke2004}). The results of this
investigation are presented in \citet{marengo2010} (PL and search for
mid-IR color excess) and in \citealt{barmby2011} (search for
extended emission).

In this paper we want to focus on \deltacep. With a distance of
0.273$\pm$0.011~kpc \citep{benedict2002} and a fundamental mode period
of 5.37~d, \deltacep{} is the second nearest Cepheid to the Solar
System (only Polaris, a first overtone Cepheid, is
closer). \deltacep{} is also known to be part of a multiple star
system. The B7-B8 III-V main sequence star \object[HD
213317]{\dcepb{}}, long suspected to be a companion of \deltacep{}
\citep{fernie1966, worley1966, vitrichenko1969} shares proper motion
with the Cepheid \citep{hoffleit1982, benedict2002}. Located
40\arcsec{} ($\sim 11,000$~AU) south of \deltacep, \dcepb{} may itself
be a binary star \citep{benedict2002}, with a low mass companion
($\sim 1.6$~\msun, F0 V) on a circular orbit with a radius of
approximately 7~mas (1.9~AU).  The three stars are part of the moving
group Cep~OB6 \citep{dezeeuw1999}, together with another 18 members
that are possibly part of an old OB association. The earliest spectral
type (B5~III) among the stars in the association suggests an age of
$\sim 50$~Myr, consistent with the estimated age of \deltacep{}
($t_{age} \simeq 60$~Myr according to the period-age-color relation
derived by \citealt{bono2005}, and using \deltacep{} colors from
\citealt{storm2004}). The proximity of \deltacep{} to the Sun, and the
presence of a hot companion which can serve as a ``probe'' of the
circumstellar environment of the system, make \deltacep{} the ideal
test case to search for evidence of Cepheids mass loss.

We report on the discovery of extended circumstellar emission around
\deltacep{} and its companions, detected at IRAC and MIPS
wavelengths. This emission appears to be physically associated with
\deltacep, and may be evidence of active mass loss from this star. We
detect a 70~\micron{} arched structure with its axis aligned with the
relative proper motion of \deltacep{} with respect to the interstellar
medium (ISM), and diffuse and filamentary emission at 5.8, 8.0 and
24~\micron. In section~\ref{sec-obs} we present our \spitzer{}
observations, while in section~\ref{sec-nebula}, \ref{sec-bowshock}
and \ref{sec-zoom} we discuss a possible explanation for the origin
and the morphology of the observed structures, and estimate the mass
loss rate required for their formation. In section~\ref{sec-summ} we
summarize our results and conclusions.

\section{Observations and Data Reduction}\label{sec-obs}

\deltacep{} was observed with IRAC and MIPS in Cycle-3, as part of our
General Observer program with PID~30666, and in Cycle-5 (IRAC only) as
part of our Guarantee Time Observer program PID~50350. The GTO
observations were meant to provide a second photometric epoch, and
deeper observations aimed at being sensitive to faint circumstellar
emission. The observations were executed on 2006 August 10 (JD =
2453957.601) and 2008 September 23 (JD = 2454732.832), using the IRAC
full frame mode. The 2006 observations were obtained with a 5 point
Gaussian dither pattern in IRAC ``stellar mode'', with total
integration times of 1.0~sec (3.6 and 4.5~\micron) and 6.0~sec (5.8
and 8.0~\micron). The 2008 data were instead acquired with a 36 point
Reuleaux dither pattern with individual exposures of 1.6~sec
integration time each, for a total exposure of 43.2~sec in each
band. The MIPS observations were obtained on 2007 July 19 (JD =
2453935.518) in the 24 and 70~\micron{} bands with the Photometry
Astronomical Observation Template. The total on-source time was
48.2~sec and 37.7~sec at 24~\micron{} and 70~\micron{} respectively.

\subsection{IRAC images and photometry}\label{ssec-irac}

The IRAC data were reduced starting from the Basic Calibrated Data
(BCDs) generated by the \spitzer{} pipeline versions S14.4.0 and
S18.7.0 (2006 and 2008 datasets respectively). Mosaic images with a
pixel scale of 0.8627\arcsec/pix were created using the IRACproc
post-BCD software \citep{schuster2006}.

In all IRAC images \deltacep{} is saturated and we have determined its
photometry using the PSF-fitting technique described in
\citet{marengo2010}. The same technique was used to estimate the
photometry of \dcepb{} in the 2008 dataset, because of column pulldown
(3.6 and 4.5~\micron) and banding (5.8 and 8.0~\micron) from the
primary star overlapping with the position of the companion, making it
impossible to derive accurate aperture photometry. For the 2006 data
the \dcepb{} brightness was determined with standard aperture
photometry. The IRAC magnitudes so derived are listed in
Table~\ref{tab:obs}.

The IRAC images are shown in Figure~\ref{fig:delta_ceph_imgs}, with
\deltacep{} and \dcepb{} PSF-subtracted. For the first time, these
stars are found to be surrounded by extended infrared emission,
brightest at 8.0~\micron{} and marginally detected at
5.8~\micron{}. The emission is diffuse, mostly contained within the
dashed arc plotted in the figure, and brighter in the region between
the two stars. The extended infrared emission is on much larger
spatial scale (up to $\sim 1.3$\arcmin, or $\sim 2.1 \cdot 10^4$~AU
at the \deltacep{} distance) than the circumstellar near-infrared
shell detected in the interferometric observations by
\citet{merand2006} with a $\sim 2$~R$_*$ $\simeq 0.4$~AU radius.

The surface brightness of the extended emission, measured in the
deeper second epoch in the four boxes shown in
Figure~\ref{fig:delta_ceph_imgs}, is listed in Table~\ref{tab:sb}. The
sky background level has been measured in the two dashed boxes also
shown in Figure~\ref{fig:delta_ceph_imgs}. To take into account the
background offset induced by array muxstriping, we have used sky box
\emph{s1} for apertures 1 and 2, and sky box \emph{s2} for apertures 3
and 4. We have corrected the IRAC surface brightness with their
extended source aperture corrections of 0.91, 0.94, 0.70 and 0.74 at
3.6, 4.5, 5.8 and 8.0~\micron{} respectively. Where no extended
emission was detected, we listed the 3$\sigma$ limits.

\subsection{MIPS images and photometry}\label{ssec-mips}

\begin{figure}
  \centering
  \includegraphics[width=0.70\textwidth, angle=0]{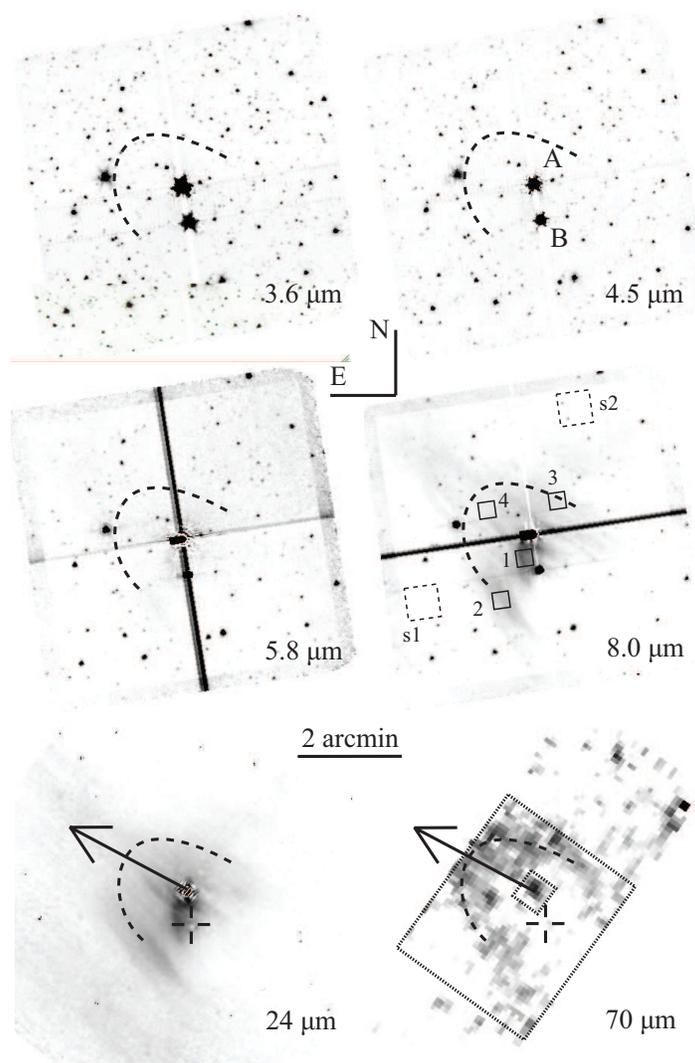}
  \caption{IRAC (2008 September 23) and MIPS (2007 July 19) images of
    \deltacep{} and surrounding area. All images are shown in a linear
    surface brightness scale ranging from 0 to 8 MJy sr$^{-1}$. Both
    \deltacep{} (A) and \dcepb{} (B) have been PSF subtracted in all
    images except MIPS 70~\micron: their location is identified either
    by the core PSF subtraction residuals, or by a cross. All
    background stars have been subtracted in the 24~\micron{}
    image. The large arrow shows the \deltacep{} proper motion
    direction relative to its local ISM. Diffuse emission surface
    brightness was measured in boxes 1 to 4, with sky level estimated
    from boxes \emph{s1} and \emph{s2}. The total 70~\micron{} flux
    within the arched structure was derived in the dotted rectangular
    aperture (large box, excluding the small box centered on the
    \deltacep{} location) plotted in the last panel. The dark and
    clear cross-bars, centered on \deltacep{} in the IRAC images, are
    column pulldown and banding artifacts.}\label{fig:delta_ceph_imgs}
\end{figure}

\begin{figure}
  \centering
  \includegraphics[width=0.85\textwidth, angle=0]{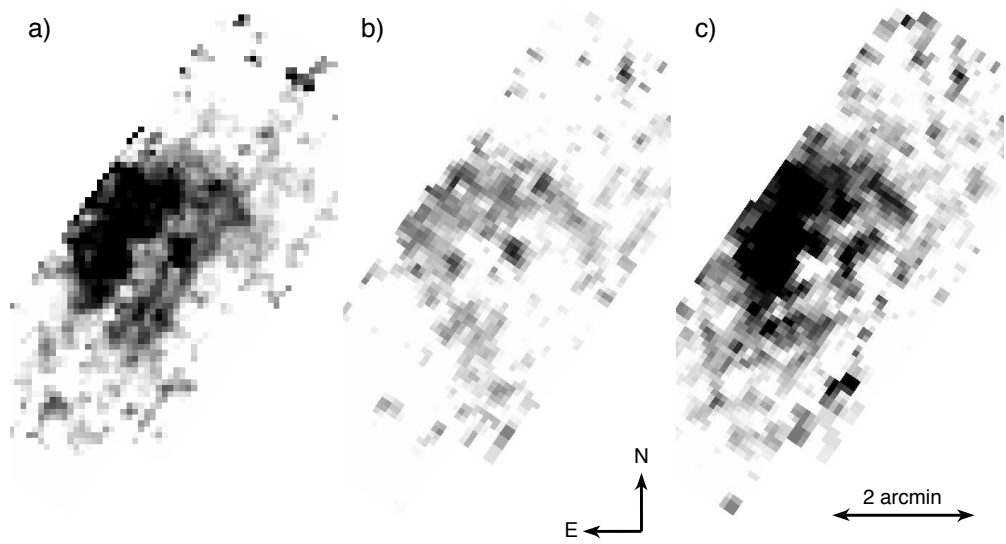}
  \caption{MIPS 70~\micron{} \deltacep{} image processed with the DAT
    pipline: \emph{(a)} no filtering (retaining strip pattern noise);
    \emph{(b)} default filtering (as in post-BCD data); \emph{(c)}
    custom filtering. All images are shown with the same linear color
    scale, from 0 to 8~MJy~sr$^{-1}$.}\label{fig:70um}
\end{figure}

We reduced the MIPS data using the MIPS instrument team Data Analysis
Tool \citep{gordon2005}. The processing of the 24~\micron{} data was
straightforward, and the final image is shown in
Figure~\ref{fig:delta_ceph_imgs}. To remove column-like instrument
artifacts, the 70~\micron{} data needs to be time filtered with the
source region excluded: this operation is complicated in our images
because of the narrow field of view, and extended luminosity covering
a substantial fraction of the detector. In Figure~\ref{fig:70um} we
show the resulting filtered images using three different filtering
choices. Panel \emph{a} shows the mosaic without any filtering applied
(leaving the strip pattern noise). The center and right panels show
instead the result of excluding the default source point region size
(\emph{b}, similar to the default region chosen by the \spitzer{}
Science Center post-BCD pipeline) and excluding the whole nebulosity
area (panel \emph{c}).

The 24~\micron{} nebulosity is morphologically similar to the one in
the IRAC 8.0~\micron{} map, and is also mostly contained within the
arched envelope plotted in Figure~\ref{fig:delta_ceph_imgs}, even
though there are faint extended lanes at larger spatial scales. It is
also brightest close to \deltacep{} and its companion, showing the
same structures detected at 8.0~\micron{}.

The 70~\micron{} image shows strong extended emission on the N-E side
of \deltacep{}, stronger along the arched envelope marked in
Figure~\ref{fig:delta_ceph_imgs}. The detailed morphology of this
structure depends on the choice for the filtering exclusion region.
The default choice highlights the high spatial frequencies allowing us
to follow the arched structure through the whole width of the array.
The larger exclusion region instead enhances the low spatial
frequencies, showing better the faint nebulosity in between the star
and the cusps of the arc. The actual shape of the nebulosity is most
likely a combination of the two, but a 70~\micron{} image with larger
field of view would be required to properly filter the map and assess
its background emission level.

The 24~\micron{} and 70~\micron{} magnitudes of \deltacep{} and its
companion, derived with standard aperture photometry as described in
\citet{marengo2010}, are listed in Table~\ref{tab:obs}. At
70~\micron{} \deltacep{} is detected with less than 3$\sigma$
significance ($93 \pm 50.5$~mJy, from the default filtered image,
which best preserves the point source). Within the uncertainty, the
color of the star (with respect to the simultaneous 24~\micron{} band)
is zero, indicating absence of significant amount of warm and cold
dust near the star (confirming the results given in
\citealt{marengo2010}).

We have estimated the 24 and 70~\micron{} surface brightness of the
extended emission in the same four boxes shown in
Figure~\ref{fig:delta_ceph_imgs}. The background level was estimated
by averaging the surface brightness values in the two sky boxes, also
marked in Figure~\ref{fig:delta_ceph_imgs}. The surface brightness is
listed in Table~\ref{tab:sb}. For the 70~\micron{} emission, we list
the two values corresponding to the different filtering options
described above.

\section{The \deltacep{} Nebula}\label{sec-nebula}

With a Galactic latitude of $+0.53^\circ$, \deltacep{} is directly
projected against the outer Galactic plane, in an area with diffuse
infrared-emitting background ``cirrus''. Inspection of wide field 24
and 70~\micron{} part of the ``\spitzer{} Mapping of the Outer
Galaxy'' survey (SMOG), from the \emph{Spitzer} archive, shows that
the diffuse emission centered around \deltacep{} is contiguous to a
system of larger Galactic structures. Understanding if the nebulosity
around \deltacep{} is physically associated with  the star, rather than a
chance superposition with background interstellar clouds, is critical
to investigate the mass loss history of the star. As part of the
\spitzer{} Cycle-5 program PID~50346, we proposed to obtain a Infrared
Spectrograph (IRS, \citealt{houck2004}) spectral energy distribution
of the \deltacep{} extended emission in key areas around the star, to
investigate the presence of diagnostic features within the IRAC and
MIPS bands, and thus identify the nature of the circumstellar
matter. The exhaustion of the spacecraft LHe refrigerant at the end of
the cryogenic mission, however, prevented this part of the program
from being executed. Program PID~40968 on 2007 December 10 obtained an
IRS high resolution spectrum centered on \deltacep{}. The data however
do not show other emission than the strong photospheric flux.

In absence of spectra for the extended emission, we compared our
\deltacep{} IRAC and MIPS images with the observations of a similar
source, for which IRS spectra are available. Following
\citet{kervella2009}, we adopted the dusty reflection nebula NGC~7023
as an ``analog'' for a Cepheid precursor surrounded by circumstellar
nebulosity. The central source in this system, the binary star
HD~200775, has a B2Ve composite spectral type, similar to the
\deltacep{} B7-8 companion \dcepb{}. With masses of 11 and 9~\msun{}
\citep{alecian2008}, the two stars at the center of NGC~7023 may
become Cepheids themselves within a few tens of million
years. Studying the diffuse emission in NGC~7023 thus offers the
chance to probe a system both representative of an early version of
\deltacep{}, and powered by a central stellar engine similar to the
\deltacep{} companion.

NGC~7023 was observed with all \spitzer{} instruments as part of the
Early Release Observation programs 717 and 1093. The central stars are
embedded in a reflection nebula characterized by bright filaments
surrounding an hourglass-shaped cavity (filled with CO emission)
probably formed by outflow activity in the HD~200775 past. The IRS
spectral energy distribution in a location along one of the bright
filaments has been presented and discussed in \citet{werner2004b}. We
have derived the surface brightness of the filaments in all IRAC and
MIPS images in the same filament location (position ``A'' in Figure~2
of \citealt{werner2004b}), and listed them in Table~\ref{tab:sb}. This
location is situated at a distance of $\sim 2.0 \cdot 10^4$~AU from
the central stars (using a distance of 430~pc for NGC~7023, according
to \citealt{ancker1997}), similar to the radius of the bright
70~\micron{} arc in the \deltacep{} system. The availability of the
complete infrared spectrum at this location allows us to determine the
prevalent mechanisms of emission responsible for the NGC~7023 diffuse
luminosity in the IRAC and MIPS bands and, by extension, provide
diagnostic tools for the unknown emission around \deltacep.

\citet{werner2004b} showed that the bulk of the flux in the IRAC bands
at 5.8 and 8.0~\micron{} is due to strong PAH and other aromatic
emission features. Fainter emission detected at 3.6 and 4.5~\micron{}
is likely related to weaker H$_2$ lines. The stronger flux in the MIPS
bands, in contrast, is associated with thermal emission from Very
Small dust Grains (VSGs) that are stochastically heated by single UV
photons generated by the central B stars. This interpretation is
confirmed by comparing the $S_{8.0}/S_{24}$ and $S_{24}/S_{70}$
surface brightness ratio of the filament with models of interstellar
dust computed by \citet{draine2007}. These models cannot be strictly
applied because of the different assumption in the spectrum of the
input energy field: two B stars in the case of the NGC~7023 nebula and
a scaled interstellar radiation field derived for the solar
neighborhood by \citet{mathis1983} for the \citet{draine2007}
models. It is nevertheless instructive to note that the flux ratios in
the NGC~7023 filament can be reproduced by assuming an irradiation of
the nebular material by a field equal to $\approx 100$ times the
interstellar radiation field in the solar neighborhood, and a PAH dust
fraction $q \sim 3$\%.

\begin{figure}
  \centering
  \includegraphics[width=0.65\textwidth, angle=-90]{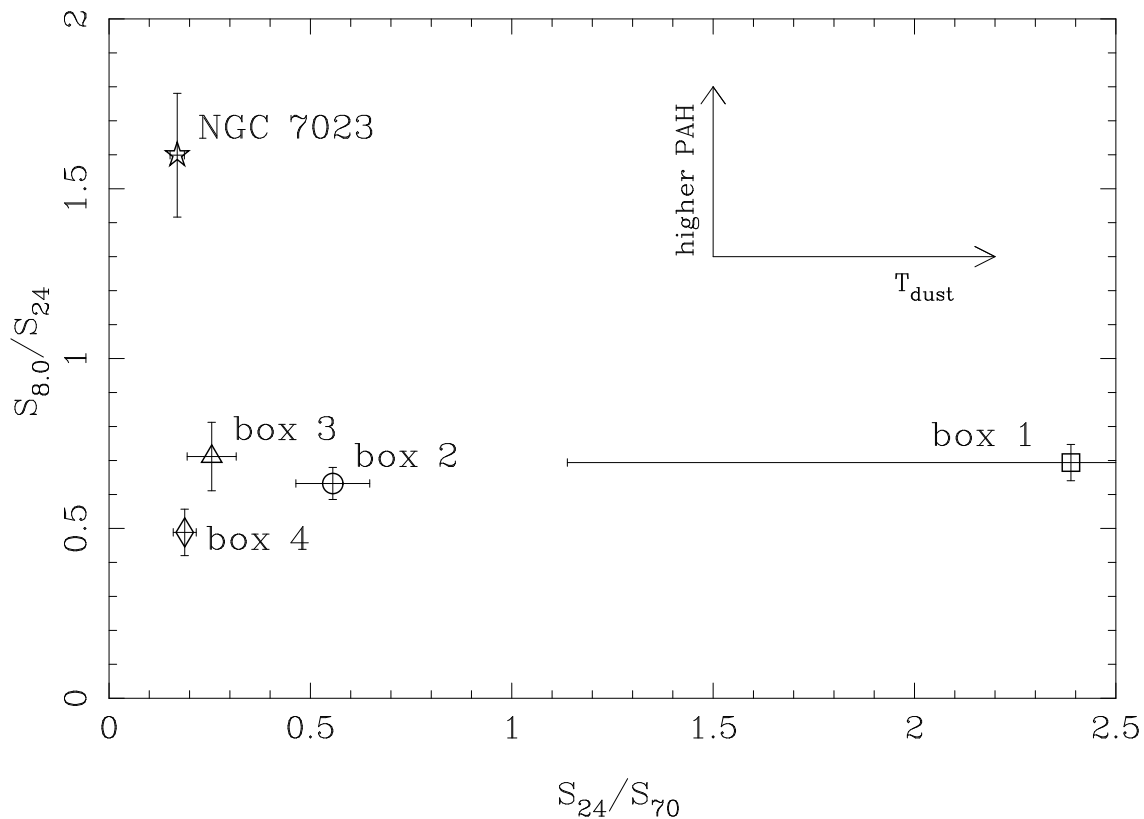}
 \caption{Flux ratios of selected regions of the extended emission
    around \deltacep{} and on one bright filament in the reflection
    nebula NGC~7023. This figure makes use of the average values of the
    70~\micron{} surface brightness listed in Table~\ref{tab:sb}, with
    their average errors. The $S_{8.0}/S_{24}$ flux intensity ratio is a
    measure of the PAH fraction in the circumstellar material, while
    the $S_{24}/S_{70}$ reflects the dust temperature, driven by the
    stellar flux stochastically heating the
    grains.}\label{fig:sb}
\end{figure}

The surface brightness ratios of the NGC~7023 filament, and of the
four locations in the \deltacep{} nebula listed in Table~\ref{tab:sb}
are plotted in Figure~\ref{fig:sb}. The figure shows that the
\deltacep{} emission has a $S_{8.0}/S_{24}$ significantly smaller than
NGC~7023, suggesting a PAH content $q \la 1$\% (based on
\citealt{draine2007}). The \deltacep{} nebula $S_{24}/S_{70}$ ratio
along the 70~\micron{} arc is as high as $\sim 3$ times the
corresponding ratio for the NGC~7023 filament, suggesting a
temperature distribution of the dust grains around the Cepheid peaked
at higher values. It is worth noting that the $S_{24}/S_{70}$ flux
ratio (and hence the dust temperature) appears to be loosely
correlated with the distance of each box from the two stars in the
\deltacep{} system. In particular, the flux ratio in box~1, located on
the bright filament in between the two stars, is more than 4 times
larger than in the other three boxes: this is strong evidence
supporting the hypothesis that the main energy input for the
\deltacep{} nebula is the two central stars, and thus that the nebula
itself is local, and not a background cloud in a chance alignment. We
do not detect any emission at 3.6 and 4.5~\micron. Nevertheless, we
cannot exclude the presence of shocked H$_2$ molecules around
\deltacep, since H$_2$ line emission would be undetected in our images
if its $S_{3.6}/S_{8.0}$ and $S_{4.5}/S_{8.0}$ ratios were similar to
what we have measured in the NGC~7023 filament.

\section{A far-IR Bow Shock?}\label{sec-bowshock}

The envelope of the 70~\micron{} emission is shaped like a paraboloid
centered on \deltacep{} and its companion. Structures of this kind are
often associated with bow shocks that form due to the interaction of a
stellar wind with the surrounding ISM (see
e.g. \citealt{wareing2006}). Bow shocks can be bright in the thermal
infrared, when the accumulated interstellar material, or the stellar
wind itself, is rich in dust. Examples of these structures around
evolved stars abound, including the case of the Asymptotic Giant
Branch (AGB) star R~Hya \citep{ueta2006, wareing2006} and the
supergiant $\alpha$~Ori \citep{ueta2008a}. The Mira variable class
prototype $o$~Ceti also shows a complex infrared bow shock structure,
associated with a UV and infrared-bright ``cometary tail'' that is
generated as the star plows at high velocity through its local ISM
\citep{ueta2008b}. If the far-IR structure surrounding \deltacep{} is
indeed a bow shock, its axis is expected to be aligned with the
direction of the star's motion through the ISM.

Following the prescription of \citet{johnson1987}, we have computed
the components of the Galactic peculiar space motion of the two stars,
$(U,V,W)_{\rm pec}$. For these calculations we assumed the following
parameters for \deltacep{} (\dcepb{}): heliocentric radial velocity
$-16.8$~\kms\ ($-21.0$~\kms) from \citet{wilson1953}; proper motion in
right ascension 16.47~mas yr$^{-1}$ (16.48~mas yr$^{-1}$) and proper
motion in declination 3.55~mas yr$^{-1}$ (4.70~mas yr$^{-1}$) from the
Hipparcos database \citep{perryman1997}. We used the parallax of
3.66~mas (3.65~mas) from \citet{benedict2002}. With the exception of
the radial velocity, the values for the two stars are all consistent
to within uncertainties. Correction for the solar motion using the
constants of \citet{dehnen1998} yields $(U,V,W)_{\rm pec}=(-6, -16,
0)$~\kms\ for \deltacep{} and $(U,V,W)_{\rm pec}=(-5, -21, 1)$~\kms\
for \dcepb{}. Projecting back into an equatorial reference frame then
gives $(V_{r},\alpha,\delta)_{\rm pec} = (-14, 8, 5)$~\kms\ and
$(V_{r},\alpha,\delta)_{\rm pec} = (-18, 8, 6)$~\kms,
respectively. This implies a space velocity for \deltacep{} of $V_{\rm
  space}\approx 17$~\kms\ along a position angle of 59$^{\circ}$ and a
space velocity for \dcepb{} of $V_{\rm space}\approx 21$~\kms\ along a
position angle of 52$^{\circ}$.

Figure~\ref{fig:delta_ceph_imgs} shows that the velocity vector of
\deltacep{} with respect to the motion of its local ISM, is roughly
aligned with the symmetry axis of the 70~\micron{} paraboloid
emission. While this is not a proof that the shape is indeed the
result of a bow shock, it strongly suggests that this may be the
case. If a bow shock forms in front of \deltacep{}, conservation of
momentum implies that the star must be losing mass, in the form of a
wind interacting with the ISM.

How much dust mass is associated with the 70~\micron{} emission? If we
assume that the dust is optically thin, we can estimate the total mass
of the dust responsible for the 70~\micron{} emission, following
\citet{evans2003}: 

\begin{equation}\label{eq:md}
\frac{M_d}{M_\odot} = 4.97 \times 10^{-14} \, f_\nu \,
\frac{D^2}{\kappa_\nu B_\nu(T_d)}
\end{equation}

\noindent
where $f_\nu$ is the total flux of the extended emission at
70~\micron{} in Jy, $D$ is the distance in kpc, $\kappa_\nu =
56$~cm$^2$~g$^{-1}$ is the dust opacity from \citet{ossenkopf1994} for
a Mathis-Rumpl-Nordsieck dust distribution \citep{mathis1977},
$B_\nu(T_d)$ is the Planck function in c.g.s units and $T_d$ is the
temperature of the dust.

Even though, as noted in Section~\ref{sec-nebula}, the small dust
grains responsible for the far-IR emission are most likely
stochastically heated, we can still estimate the mode of the grain
temperature distribution as the color temperature of the
$S_{24}/S_{70}$ flux ratio. Boxes 2, 3 and 4, lying on the
70~\micron{} bright arc, have color temperature in the range of 85 --
110~K, comparable with the color temperature of the NGC~7023 filament
($\sim 80$~K). The color temperature of the dust in box 1, placed in
between \deltacep{} and \dcepb{}, is however much higher ($\sim
180$~K), as expected if heated by the radiation from the two close-by
stars.

The total 70~\micron{} flux density of the arched emission, measured
in the dotted rectangular annulus in Figure~\ref{fig:delta_ceph_imgs}
is between 1.0 and 2.8~Jy, depending on which data reduction product
is used. We used a rectangular aperture to exclude the flux from the
marginally detected star. From Equation~\ref{eq:md}, using the average
color temperature $T_d \approx 100$~K along the arched structure, and
$f_\nu \approx 2$~Jy, we have $M_d \approx 6 \cdot
10^{-7}$~M$_\odot$. If we assume a gas to dust mass ratio of $\sim
100$, typical of ISM and circumstellar dust, we obtain a total mass of
gas and dust of $\approx 6 \cdot 10^{-5}$~M$_\odot$. A minimum age of
this structure is given by the time required by a stellar wind to
reach the arc from \deltacep{} (or its companion). In the case of
\deltacep{}, the wind speed is expected to be comparable to the escape
velocity ($\sim 100$~km~s$^{-1}$; e.g. \citealt{welch1988}),
i.e. several times higher than is typical for AGB stars. This implies
a minimum age of $\ga 10^3$~yr for the structure, and a maximum mass
loss rate of $\dot M \approx 6 \cdot
10^{-8}$~M$_\odot$~yr$^{-1}$. These value is at the upper end of the
Cepheids mass loss rates proposed by \citet{deasy1988}.

We can also derive a minimum mass loss rate required to produce the
observed bow shock structure using the standard stellar wind/ram
pressure balance condition (e.g., \citealt{raga2008}):

\begin{equation}\label{eq:bs}
\dot{M} = \frac{4\pi d^{2}_{s}\rho_{0}v_{\star}^2}{v_{w}}
\end{equation}

\noindent

Here $d_{s}$ is the standoff distance of the bow shock, $\rho_{0}$ is
the ambient ISM density, $v_{\star}$ is the space velocity of the star
relative to the ambient medium, and $v_{w}$ is the wind outflow
velocity. From our {\it Spitzer} images we measure $d_{s}\approx
1.3$~arcmin (0.103~pc). We assume that the surrounding medium is pure
atomic hydrogen and, from \citet{dickey1990}, a typical particle
density in the Galactic plane is $\sim 0.55$~cm$^{-3}$, implying
$\rho_{0} \approx 9.2 \cdot 10^{-25}$~g~cm$^{-3}$. Finally, adopting
an expected outflow velocity for the Cepheid wind of $v_{w} \approx
100$~km~s$^{-1}$ (see above), we find $\dot M \approx 5 \cdot
10^{-9}$~M$_{\odot}$~yr$^{-1}$.  While this value can be considered
only approximate given the uncertainties in the wind speed for the
star and the local ISM density, it is nonetheless consistent with the
predictions of \citet{deasy1988} for Cepheid mass-loss rates (albeit
significantly higher than the $\dot M_{puls} \simeq 3.6 \cdot
10^{-10}$~M$_\odot$~yr$^{-1}$ predicted by \citealt{neilson2008} for
this star). In particular, if \deltacep{} sits in a over-dense area of
the Galactic plane, as the SMOG and IRAS maps imply, the ISM density
(and thus the minimum $\dot M$) could be higher.

The fact that the mass-loss rate derived using this approach is one
magnitude lower than the estimate computed above, using the observed
properties of the dust emission, suggests that part of the bow shock
may be comprised of swept up interstellar material rather than
entirely shed from the star. Our images in fact do not allow us to
determine how much of the detected mass is coming from the star,
rather than being swept from the ISM by the stellar wind (see also
discussion below on why at least the PAH emission detected at
8.0~\micron{} is likely of interstellar origin). The presence of a
bow-shock structure, however, argues in favor of a stellar wind
pushing against the ISM, and thus at least part of this mass ought to
come from the star, with a mass loss rate of at least $5 \cdot
10^{-9}$~M$_{\odot}$~yr$^{-1}$ and up to $6 \cdot
10^{-8}$~M$_{\odot}$~yr$^{-1}$.

Given the high luminosity of the Cepheid and its companion, it is fair
to ask if the observed dust-free bubble around the stars may be
generated by effect of radiation pressure alone, without the need of
invoking a Cepheid winds. The radiation pressure has an inverse square
law variation and therefore acts like a repulsive ``Coulomb''
force. If an interstellar grain of mass $m$ has speed $v =
17$~km~s$^{-1}$ relative to the star, then the distance of closest
approach $D$ is given by:

\begin{equation}\label{eq:coulomb}
\frac{1}{2} m v^2 = \frac{L_*}{4 \pi D c} \pi a^2 Q
\end{equation}

\noindent
where $L_*$ is the luminosity of the Cepheid ($\sim 2,000$~L$_\odot$),
$c$ is the speed of light, $a$ is the radius of a grain and $Q$ is a
dimensionless number to express the effective momentum coupling to the
geometric cross section. For a spherical grain of radius $a$ and
density $\rho_s \simeq 3$~g~cm$^{-3}$ (typical value of astronomical
silicates):

\begin{equation}\label{eq:min-approach}
D = \frac{3}{8 \pi a \rho_s} \frac{Q L_*}{v^2 c}
\end{equation}

Given that dust grains are opaque at optical and near-IR radiation, we
can assume $Q = 1$. According to equation~\ref{eq:min-approach} a
distance of closest approach $D \simeq 0.1$~pc is obtained for dust
grains with radius $a \simeq 0.11$~\micron.

However, based on the discussion in Section~\ref{sec-nebula}, the 24
and 70~\micron{} emission is most likely due to stochastically heated
VSGs, that have a typical radius of $\sim 10^{-3}$~\micron. Such
grains would be blown out by radiation pressure at a distance two
order of magnitudes larger than the observed $D \simeq
0.103$~pc. Furthermore, the opening angle of a dust-free bubble
created by radiation pressure would be much larger than the one
observed: for the direction orthogonal to the space velocity of the
star with respect to the ISM, the dust grains will have a zero radial
component velocity, and will be blown out at a distance much larger
than $D$ derived from Equation~\ref{eq:min-approach}, in contrast to
the narrow arc observed in Figure~\ref{fig:delta_ceph_imgs}. The small
size of the emitting grains, and the shape of the arched structure,
imply that gas drag is likely to play a strong role in preventing the
dust grains from being blown out by the radiation pressure. While
radiation pressure is likely to play some role in the dynamic of the
system, we thus still favor the hypothesis that the arched structure
is due to a mechanical force acting on both interstellar gas and dust,
due to a wind originating from the stars in the system.

Detection of emission lines from the shocked material along the arc
would confirm that the structure is indeed a bow shock generated by a
stellar wind pushing on the ISM. The power $P$ in the stellar wind is
given by $\frac{1}{2} \dot M v_w^2$. With $v_w \simeq 100$~km~s$^{-1}$
and a mass loss rate in the range of $5 \cdot 10^{-9}$ to $6 \cdot
10^{-8}$~M$_\odot$~yr$^{-1}$, $P \simeq 4 \cdot 10^{-3}$ to $5 \cdot
10^{-2}$~L$_\odot$. This energy will likely be dissipated in a
radiative shock, and emission lines (among which H$_2$ or CO) could
be observable. This was one of the main motivations for our
\emph{Spitzer}/IRS cycle-5 program, that was unfortunately not
executed before the end of the cryogenic mission. We have plans to
attempt the detection of these lines with ground-based near-IR
observations (searching for the $\sim 2$~\micron{} H$_2$ lines) and
possibly with the Herschel Space Telescope as a follow-up of our OT-1
imaging proposal. L Band VLA 21~cm observations could also be
attempted in search of H\textsc{I} assuming that any structured
emission around the star can be successfully disentangled from
Galactic foreground and background emission.

\section{The Nebulosity Near the Two Stars}\label{sec-zoom}

\begin{figure}
  \centering
  \includegraphics[width=0.85\textwidth, angle=0]{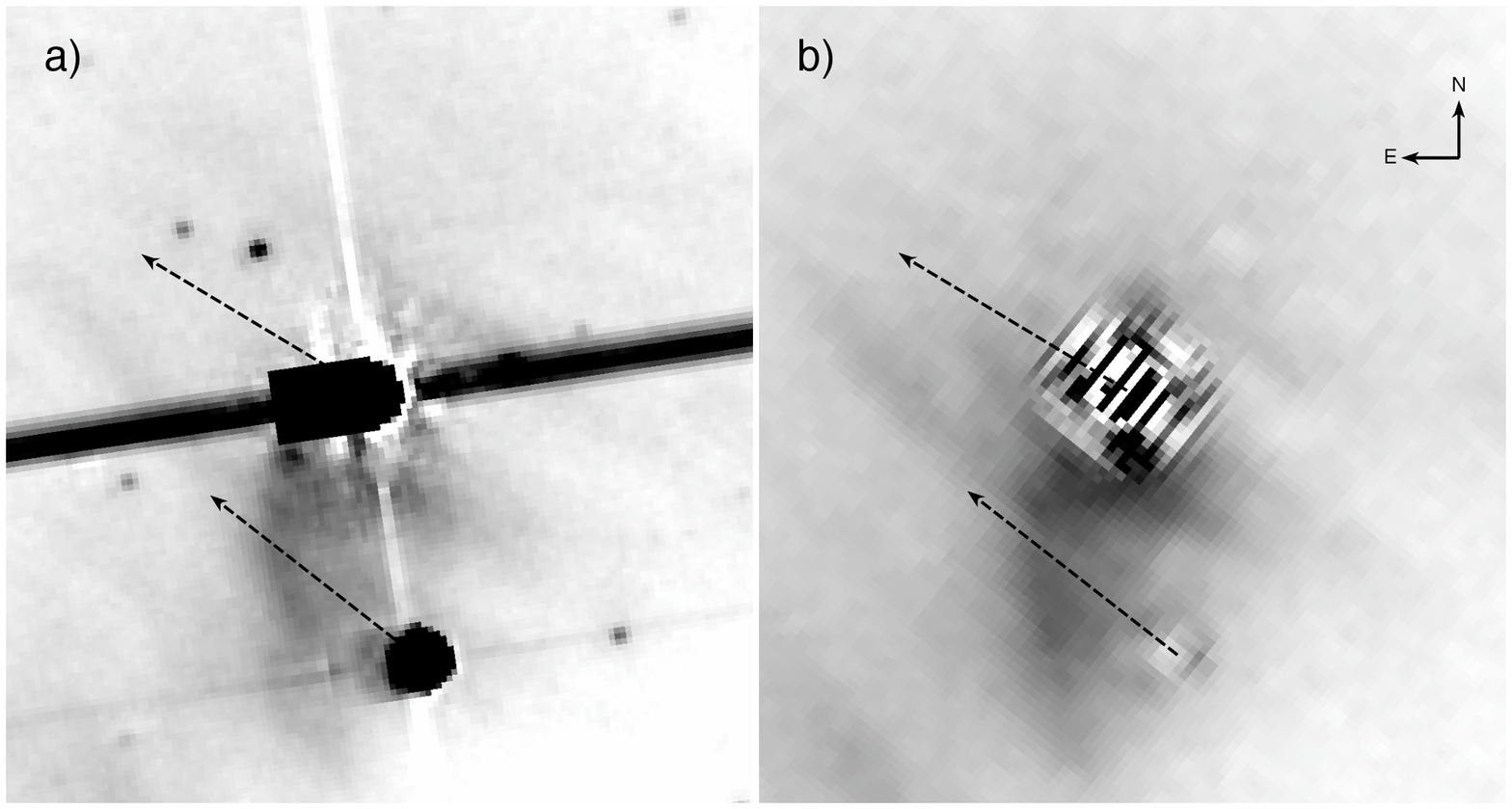}
  \caption{Details of the IRAC 8.0~\micron{} (\emph{a}) and MIPS
    24~\micron{} (\emph{b}) images showing the area in proximity of
    \deltacep{} and \dcepb{}. Both stars have been PSF subtracted in
    the two images: \deltacep{} is the star north of \dcepb{}. All
    background stars have been removed in the 24~\micron{} image to
    highlight the diffuse emission. The images are shown in linear
    color scale from 0 to 8 MJy/sr surface brightness. The two arrows,
    having a length of 40~arcsec, point to the direction of the
    relative motion of the two stars with respect to the
    ISM.}\label{fig:zoom}
\end{figure}

Figure~\ref{fig:zoom} shows the 8.0 and 24~\micron{} emission in the
region immediately around \deltacep{} and \dcepb{}. This is the area
where the diffuse circumstellar emission is stronger at these two
wavelength (box 1 in Figure~\ref{fig:delta_ceph_imgs}), and lowest at
70~\micron{}, corresponding to a color temperature of $\sim
180$~K. The figure shows the rather complex morphology of this
emission, with a small arched structure that seems to be connecting
the two stars, and other filaments arching away from \deltacep{}. Can
the morphology of this emission be explained in context of the
\deltacep{} mass loss hypothesis? In absence of a complete geometry of
the region (in particular, the relative position of the two stars
along the line of sight) and kinematic data for the nebulosity, we can
only speculate about the origin of these structures.

One possibility is that the arched structure is in fact a second bow
shock, this time associated to \dcepb{}. While the structure is not
perfectly aligned with the relative motion of \dcepb{} with respect to
the ISM (but still within the uncertainties), it is instructive to
evaluate what would be the mass loss rate required for its
creation. The standoff distance of the arc from \dcepb{} is in this
case $d_{s} \sim 26$~arcsec (0.034~pc), and the escape velocity from
\dcepb{} (a B7-B8 main sequence star) is $\approx
500$~km~s$^{-1}$. Using equation~\ref{eq:bs}, one obtains a required
mass loss rate of $\approx 10^{-10}$~M$_\odot$~yr$^{-1}$, more than
one order of magnitude smaller than the minimum mass loss rate that we
have estimated for \deltacep{}. This is in agreement with the
expectation that a late-B star like \dcepb{} should not have a strong
wind (see e.g. \citealt{hempel2003}), as winds drop off strongly in
late B stars (\citealt{kudritzki2000}; see also eq.~25 in
\citealt{vink2001} predicting $\log_{10} \dot M \simeq -11.6$ for a B7
main sequence star with $T_{eff} \simeq 12,500$~K, $L \simeq
160$~L$_\odot$, $M \simeq 4$~M$_\odot$, solar metallicity and a
galactic value of $v_\infty/v_{esc} = 1.3$). The imperfect alignment
of the arc with respect to the \dcepb{} motion, and the brightening of
the arc near \deltacep{} could be explained by the interaction of the
two stellar winds and with the irradiation of the \dcepb{} bow shock
from the Cepheid photons. This explanation requires that the distance
of the two stars along the line of sight is significantly larger than
their distance on the plane of the sky, outside the 3-dimensional
paraboloid partially evacuated by the \deltacep{} wind. Conversely, if
the two stars were on the same plane, a bow shock from \dcepb{} would
be blown away by a stronger wind from \deltacep{}. The uncertainly in
the parallax of the two stars is large enough to allow this
geometry. If this hypothesis is correct, then the possibility that
\deltacep{} is the origin of a wind responsible for the large,
70~\micron{} bright, arc is strongly supported, as the weak wind from
\dcepb{} would not be sufficient to generate a bow shock at the
standoff distance of 0.103~pc.

Another possibility involves wind mass transfer between \deltacep{}
and \dcepb{}, similar to the structures observed between $o$~Ceti and
its compact companion in the UV \citep{karovska1997} but also in the
thermal infrared \citep{marengo2001}. The much larger separation of
\deltacep{} from \dcepb{}, with respect to the separation of $o$~Ceti
from its companion, makes however this interpretation very
unlikely. An asymmetric mass ejection from \deltacep{}, in the chance
direction of \dcepb{} is also very unlikely because it would require
non-radial oscillations with such strength that has never been observed
in a Cepheid, or strong irradiation of the \deltacep{} atmosphere by
the \dcepb{} photons which is however negligible. The possibility that
the arched structure is in fact originating from the F0~V companion of
\dcepb{}, being irradiated by the nearby B star, is also to be
discounted. While the flux from \dcepb{} contributes to as much as
$\sim 10$\% to the photospheric flux of the F0~V companion, a plume
from this star (orbiting \dcepb{} with a period of $\sim 1$~yr, see
\citealt{benedict2002}) would generate a tightly wound
spiral. Assuming again a wind from \dcepb{} as fast as its escape
velocity, the spacing in the plume spiral would be $\approx 100$~AU,
much smaller than the wide arc observed in the images.

The exact nature of the structures seen in proximity of the two stars
is difficult to assess with the available data. However, the presence
of enhanced nebulosity nearest to the two stars is strongly supportive
of the hypothesis that the overall structures that we have detected
around \deltacep{} and its companion are local, rather than a chance
alignement with ISM background cirrus.

\section{Discussion and conclusions}\label{sec-summ}

Our \deltacep{} \emph{Spitzer} data provides a new and unique view of
the recent and current mass loss history of a Classical Cepheid
star. This system is special, in that the presence of a B spectral
type companion provides enough UV photons to excite the circumstellar
molecules, making it possible to map them at infrared wavelengths. In
the sample of 29 Cepheids observed with \spitzer{}
\citep{marengo2010}, 8 have hot companions, but \dcepb{} is
significantly hotter than all of the others, with the exception of
S~Mus that also shows signs of extended emission
\citep{barmby2011}. The question arises as to whether \deltacep{} may
be unique in having mass loss as the result of the presence of
\dcepb{}.  The answer is most likely negative, given that the wide
separation of the companion is probably too large to affect the mass
loss properties of the Cepheid. None of the \deltacep{} properties as
Cepheid (period, luminosity, or amplitude) suggest it is unusually
likely to have a mass loss episode. We can then expect the mass-loss
properties of \deltacep{} to be representative of Classical Cepheids
in general, with the only difference that the presence of \dcepb{} and
the peculiar motion of \deltacep{} with respect to the ISM makes any
stellar ejecta easier to detect.

The presence of local (i.e. circumstellar) structures resembling a bow
shock is strong evidence that one or both stars in the system are
currently losing mass. The 70~\micron{} emission is better aligned
with respect to the \deltacep{} position and relative proper motion
vector (with respect to the ISM), although the low resolution and
sensitivity at this wavelength is not good enough to rule out
alignment with the companion \dcepb{}. While the wind responsible for
this structure may be arising from either star (or both), the
circumstellar shell found by near-IR interferometers
\citep{merand2006} around \deltacep{} strongly supports the
possibility that the Cepheid star is the origin of the wind. The weak
wind expected by a late B star like \dcepb{} is also unlikely to be
able to form a bow shock with the observed standoff distance of the
70~\micron{} structure. An independent confirmation of this
expectation could be obtained by showing that the arched filament near
the two stars is indeed a secondary bow shock generated by a weak wind
coming from \dcepb{}.

\citet{kervella2009} found extensive ($\sim 100$ to $1000$~AU)
``warm'' ($\sim 100$~K) emission in Spitzer and ground-based mid-IR
images of RS~Pup and $\ell$~Car, which they attributed to dusty mass
loss from these stars (as opposed to cold dust emission, $\sim 40$~K,
in the large massive nebula around RS~Pup, attributed to ISM material
compressed by the Cepheid wind). Our images suggest that part of the
material responsible for the extended emission around \deltacep{} is
of interstellar origin. Dust grains at $\sim 80$~K (the color
temperature of the arc) are too cold to explain the observed
$S_{8.0}/S_{24}$ ratio. On the other hand, PAH emission as described
by \citet{draine2007} provides a reasonable fit of the data. Given the
typical abundances of Cepheid atmospheres (where C/O $<$ 1), if dust
were to form, its composition would largely be characterized by O-rich
silicates and not by carbon-based PAH. Furthermore, if indeed
\deltacep{} ($T_{eff} \simeq 5800$~K) is currently losing mass, the
wind is likely to be atomic and dust free, at least in proximity of
the stellar photosphere (e.g. \citealt{glassgold1983}). This is in
agreement with the absence of significant infrared excess we measured
close to the star (see \citealt{marengo2010}), and the fact that the
circumstellar shell found by \citet{merand2006} would have an
equilibrium temperature too high for astrophysical dust survival.

We can conclude that the structures we detect with \emph{Spitzer} are
likely the consequence of the interaction of a \deltacep{} wind with
the local ISM. The low dust content of this wind near the star implies
a different driving mechanism than the dust-driven wind commonly
associated to evolved stars. As suggested by \citet{merand2006}, as an
explanation for the shell structures observed with interferometry, it
is conceivable that this ``dustless'' wind from \deltacep{} is perhaps
triggered by the pulsation and shocks crossing the atmosphere. As it
expands, this wind interacts with the ISM, leading to the formation of
the bow-shock oriented with the proper motion of the Cepheid. The
dilution of the ISM by the Cepheid wind is in agreement with the
observed low abundance of PAH (less than one third of the abundance in
the NCG~7023 nebula). Without detailed dynamical data it is however
difficult to accurately determine which fraction of the dust
responsible for the strong 24 and 70~\micron{} emission is of
interstellar origin, and how much is condensed in the outflow as it
expands and cools.

A star with the evolutionary mass of \deltacep{} will stay in the
instability strip for a total time $t_{cr} \sim 1.5$~Myr (second and
third crossing of the strip during the star's blue loop, see
\citealt{bono2000}). If mass loss with a rate at the upper limit of
our measured range ($\dot M \approx 6 \cdot
10^{-8}$~M$_\odot$~yr$^{-1}$) is sustained through all the Cepheid
phase, the total mass that could be ejected would be as high as $\sim
0.1$~M$_\odot$. This is $\approx 2$\% of the total current pulsational
mass of \deltacep{}. While not enough to completely account for the
Cepheid mass discrepancy, this value may represent a significant
portion of it, with possibly more mass lost at the onset of central
He-burning (RG) phase, before the star crossed the Cepheid instability
strip. A mass loss rate closer to our lower limit ($\dot M \approx 5
\cdot 10^{-9}$~M$_\odot$~yr$^{-1}$), however, would only result in the
ejection of a more modest $\approx 0.01$~M$_\odot$, or $\approx 0.2$\%
of the current pulsational mass of \deltacep{}. Firm constraints on
the Cepheid mass loss rate will also have a significant impact on the
plausibility of the predicted Mass-Luminosity relation of
intermediate-mass stars during central helium burning phases.

Far-IR images with higher angular resolution, larger field of view and
better sensitivity are required to provide a better estimate of the
current \deltacep{} mass loss rate, and the contribution of mass loss
to solve the Cepheids mass discrepancy. Targeted observations with the
Herschel space telescope, and high sensitivity radio maps in atomic
and molecular H and CO, may be capable of characterizing the
morphology and thermal structure of the large scale emission around
this star, thus probing its mass loss history before and during the
onset of the Cepheid phase.

In summary, our Spitzer images have detected strong extended
nebulosity around \deltacep{} and its companion \dcepb{}. This
nebulosity is likely to be local, due to mass loss processes from one
or both stars. A large scale far-IR arc, in particular, may be
associated with a bow shock generated by the wind of \deltacep{} as it
moves through the ISM. This is one of the strongest direct evidence to
date of Cepheid mass loss, and the best available direct measurement
of the mass loss rate for a Classical Cepheid, which we estimate to be
in the range of $\approx 5 \cdot 10^{-9}$ to $6 \cdot
10^{-8}$~M$_\odot$~yr$^{-1}$.


\acknowledgments

This work is based on observations made with the \textit{Spitzer Space
  Telescope}, which is operated by the Jet Propulsion Laboratory,
California Institute of Technology under NASA contract 1407.
P.B. acknowledges research support through a Discovery Grant from the
Natural Sciences and Engineering Research Council of Canada. NRE
acknowledges support from the Chandra X-ray Center, NASA contract
NAS8-03060. This work was supported in part by the NSF REU and
Department of Defence ASSURE programs under grant no. 0754568.

{\it Facilities:} \facility{Spitzer (IRAC, MIPS)}.



\begin{deluxetable}{ccccccc}
\tabletypesize{\footnotesize}
\tablewidth{0pt}
\tablecaption{Observations Log and Magnitudes}
\tablehead{
  \colhead{JD-2,400,000} &
  \colhead{[3.6]} &
  \colhead{[4.5]} &
  \colhead{[5.8]} &
  \colhead{[8.0]} &
  \colhead{[24]}  &
  \colhead{[70]}}
\startdata
\sidehead{\deltacep}
53935.018\tablenotemark{a}
          &      \nodata    &      \nodata   
          &      \nodata    &      \nodata   
          & 2.120$\pm$0.001 & 2.30$\pm$0.59  \\
53957.101\tablenotemark{b}
         & 2.174$\pm$0.044 & 2.183$\pm$0.037
         & 2.166$\pm$0.032 & 2.150$\pm$0.039
         &      \nodata    &      \nodata    \\
54732.332\tablenotemark{b}
         & 2.372$\pm$0.014 & 2.392$\pm$0.015
         & 2.358$\pm$0.019 & 2.345$\pm$0.033
         &      \nodata    &      \nodata    \\
\\
\hline
\sidehead{\dcepb{}}
53935.018\tablenotemark{d}
          &      \nodata    &      \nodata   
          &      \nodata    &      \nodata   
          & 6.139$\pm$0.087 &      \nodata    \\
53957.101\tablenotemark{c}
         & 6.391$\pm$0.010 & 6.399$\pm$0.010
         & 6.413$\pm$0.006 & 6.406$\pm$0.005
         &      \nodata    &      \nodata    \\
54732.332\tablenotemark{b}
         & 6.237$\pm$0.137 & 6.383$\pm$0.117
         & 6.402$\pm$0.304 & 6.363$\pm$0.376
         &      \nodata    &      \nodata    \\
\enddata
\tablenotetext{a~}{20\arcsec{} radius aperture, with a 40\arcsec --
  50\arcsec{} sky annulus and aperture correction of 1.13 at
  24~\micron; 16\arcsec{} radius aperture with 18\arcsec --
  39\arcsec{} sky annulus and 2.04 aperture correction at 70~\micron.} 
\tablenotetext{b~}{PSF-fitting photometry as described in
 \citet{marengo2010}.} 
\tablenotetext{c~}{6.1\arcsec{} radius aperture, with
  40\arcsec--50\arcsec{} sky annulus and aperture correction of 1.048,
  1.052, 1.053 and 1.062 at 3.6, 4.5, 5.8 and 8.0~\micron{}
  respectively. The photometry has been corrected for the field
  dependent distortion, and pixel-phase dependent uncertainties were
  added at 3.6 and 4.5~\micron{} (as explained in
  \citet{marengo2010}.} 
\tablenotetext{d~}{7\arcsec{} radius aperture, with
 40\arcsec--50\arcsec{} sky annulus and 1.61 aperture correction.} 
\label{tab:obs}
\end{deluxetable}
\clearpage

\begin{deluxetable}{cccccccc}
\tabletypesize{\footnotesize}
\rotate
\tablewidth{0pt}
\tablecaption{Extended Emission Surface Brightness\tablenotemark{a}}
\tablehead{
  \colhead{Box} &
  \colhead{$S_{3.6}$} &
  \colhead{$S_{4.5}$} &
  \colhead{$S_{5.8}$} &
  \colhead{$S_{8.0}$} &
  \colhead{$S_{24}$}  &
  \colhead{$S_{70}$ (DAT default)\tablenotemark{b}} &
  \colhead{$S_{70}$ (DAT custom)\tablenotemark{c}}  \\
 \colhead{} &
  \colhead{[MJy sr$^{-1}$]} &
  \colhead{[MJy sr$^{-1}$]} &
  \colhead{[MJy sr$^{-1}$]} &
  \colhead{[MJy sr$^{-1}$]} &
  \colhead{[MJy sr$^{-1}$]} &
  \colhead{[MJy sr$^{-1}$]} &
  \colhead{[MJy sr$^{-1}$]}}
\startdata
1 & $<0.33$       & $<0.27$       & $0.82\pm0.12$ & $3.56\pm0.20$
  & $5.13\pm0.27$ & $2.38 \pm 0.72$ & $1.69 \pm 1.02$ \\
2 & $<0.18$       & $<0.18$       & $0.45\pm0.07$ & $1.60\pm0.08$
  & $2.53\pm0.14$ & $4.58 \pm 0.61$ & $4.52 \pm 0.80$ \\
3 & $<0.42$       & $<0.33$       & $<0.36$       & $0.79\pm0.08$
  & $1.11\pm0.11$ &$4.78 \pm 0.78$ & $3.77 \pm 0.92$ \\
4 & $<0.42$       & $<0.33$       & $<0.36$       & $0.63\pm0.07$
  & $1.29\pm0.11$ & $5.23 \pm 0.62$ & $8.41 \pm 0.91$ \\
\hline
NGC~7023 & $53.2\pm7.5$ & $33.0\pm3.4$ & $300.\pm24.$ & $769.\pm60.$
         & $481.\pm40.$ & $2840\pm180$ & \nodata \\
\enddata
\tablenotetext{a~}{Values without uncertainty are 3$\sigma$ upper limits.}
\tablenotetext{b~}{From the DAT reduced image, standard filtering
  (panel \emph{b} Figure~\ref{fig:70um}).}
\tablenotetext{c~}{From the DAT reduced image, custom filtering (panel
  \emph{c} Figure~\ref{fig:70um}).}
\label{tab:sb}
\end{deluxetable}
\clearpage


\newpage

\newpage

\newpage

\begin{thebibliography}{}

\bibitem[Alecian et al.(2008)]{alecian2008} Alecian, E. et al. 2008,
  \mnras, 385, 391
\bibitem[Barmby et al.(2011)]{barmby2011} Barmby, P., Marengo, M.,
  Evans, N. R., Bono, G., Huelsman, D., Su, K. Y. L., Welch, D. L. \&
  Fazio, G., G. 2011, AJ, 141, 42
\bibitem[Beaulieu et al.(2001)]{beaulieu2001} Beaulieu, J. P.,
  Buchler, J. R. \& Kollath, Z. 2001, \aap, 373, 164
\bibitem[Beichman et al.(1988)]{beichman1988} Beichman, C. A.,
  Neugerbauer, G., Habing, H. J., Clegg, P. E. \& Chester,
  T. J., eds. 1988, IRAS Catalogs and Atlases., Vol. 1: Explanatory
  Supplement (Pasadena, CA: JPL) 
\bibitem[Benedict et al.(2002)]{benedict2002} Benedict, G. F. et
  al. 2002, \aj, 124, 1695
\bibitem[Bono et al.(2000)]{bono2000} Bono, G., Caputo, F., Cassisi,
  S., Marconi, M., Piersanti, L. \& Tornamb\`e, A. 2000, \apj, 543,
  955 
\bibitem[Bono et al.(2001)]{bono2001} Bono, G., Gieren, W. Marconi,
 M., Fouqu\'e \& P. Caputo, F. 2001, \apj, 563, 319
\bibitem[Bono et al.(2002)]{bono2002} Bono, G., Castellani, V. \&
  Marconi, M. 2002, \apj, 565, L83
\bibitem[Bono et al.(2005)]{bono2005} Bono, G., Marconi, M., Cassisi,
  S., Caputo, F., Gieren, W. \& Pietrzynski, G. 2005, \apj, 621, 966
\bibitem[Caputo et al.(2005)]{caputo2005} Caputo, F., Bono, G.,
  Fiorentino, G., Marconi, M. \& Musella, I. 2005, \apj, 629, 1021
\bibitem[Castellani et al.(2006)]{castellani2006} Castellani, V.,
  Iannicola, G., Bono, G., Zoccali, M., Cassisi, S. \& Buonanno,
  R. 2006, \aap, 446, 569 
\bibitem[Christy(1968)]{christy1968} Christy, R. F. 1968,
  Quart. J. Roy. Astron. Soc, 9, 13
\bibitem[Deasy(1988)]{deasy1988} Deasy, H. P. 1988, \mnras, 231, 673
\bibitem[DeJager \& Nieuwenhuijzen(1997)]{dejager1997} de Jager C. \&
  Nieuwenhuijzen, H. 1997, \mnras, 290, L50
\bibitem[Dehnen \& Binney(1998)]{dehnen1998} Dehnen, W. \& Binney,
  J. 1998, \mnras, 298, 387 
\bibitem[de Zeeuw et al.(1999)]{dezeeuw1999} de Zeeuw, P. T.,
  Hoogerwerf, R., de Bruijne, J. H. J., Brown, A. G. A. \& Blaauw,
  A. 1999, \aj, 117, 354
\bibitem[Dickey \& Lockman(1990)]{dickey1990} Dickey, J. M. \&
  Lockman, F. J. 1990, \araa, 28, 215 
\bibitem[Draine \& Li(2007)]{draine2007} Draine, B.T. \& Li, A. 2007,
  \apj, 657, 810
\bibitem[Evans et al.(2003)]{evans2003} Evans, A. Stickel, M., van
  Loon, J. Th., Eyres, S. P. S., Hopwood, M., E., L. \& Penny,
  A. J. 2003, \aap, 408, L9
\bibitem[Evans et al.(2008)]{evans2008} Evans, N. R., Schaefer, G. H.,
  Bond, H. E., Bono, G., Karovska, M., Nelan, E., Sasselov, D. \&
  Mason, B. 2008, \aj, 136, 1137
\bibitem[Fazio et al.(2004)]{fazio2004} Fazio, G. G. et al. 2004,
  \apjs, 154, 10
\bibitem[Fernie(1966)]{fernie1966} Fernie, J. D. 1966, \aj, 71, 119 
\bibitem[Fricke et al.(1972)]{fricke1972} Fricke, K., Stobie, R. S. \&
 Strittmatter, P. A. 1972, \apj, 171, 593
\bibitem[Gehrz et al.(2007)]{gehrz2007} Gehrz, R. D. et al. 2007,
  Rev. Sci. Instrum., 78, 011302
\bibitem[Glassgold \& Huggins(1983)]{glassgold1983} Glassgold,
  A. E. \& Huggins, P. J. 1983, \mnras, 203, 517 
\bibitem[Goodricke \& Staatsoper(1786)]{goodricke1786} Goodricke,
  J. B. \& Staatsoper,Bayerische 1786, Phil. Trans. of the Royal
  Astronomical Soc., 76, 48 
\bibitem[Gordon et al.(2005)]{gordon2005} Gordon, K. D. et al. 2005,
  \pasp, 117, 503
\bibitem[Hempel \& Holweger(2003)]{hempel2003} Hempel, M. \& Holweger,
  H. 2003, \aap, 408, 1065
\bibitem[Hoffleit \& Jaschek(1982)]{hoffleit1982} Hoffleit, D. \&
  Jaschek, C. 1982, The Bright Star Catalogue (4th ed.; New Haven:
  Yale Univ. Obs.)
\bibitem[Houck et al.(2004)]{houck2004} Houck, J. R. et al. 2004,
  \apjs, 154, 18
\bibitem[Johnson \& Soderblom(1987)]{johnson1987} Johnson, D. R. H. \&
  Soderblom, D. R. 1987, \aj, 93, 864 
\bibitem[Karovska et al.(1997)]{karovska1997} Karovska, M., Hack, W.,
  Raymond, J. \& Guinan, E. 1997, \apj, 482, L175
\bibitem[Keller \& Wood(2006)]{keller2006} Keller, S. C. \& Wood,
  P. R. 2006, \apj, 642, 834
\bibitem[Kervella et al.(2006)]{kervella2006} Kervella, P., M\'erand,
  A., Perrin, G. \& Coud\'e Du Foresto, V. 2006, \aap, 448, 623
\bibitem[Kervella et al.(2008)]{kervella2008} Kervella, P., M\'erand,
  A., Szabados, L., Fouqu\'e, P., Bersier, D., Pompei, E. \& Perrin,
  G. 2008, \aap, 480, 167
\bibitem[Kervella et al.(2009)]{kervella2009} Kervella, P., M\'erand,
  A. \& Gallenne, A. 2009, \aap, 498, 425
\bibitem[Kudritzki \& Puls(2000)]{kudritzki2000} Kudritzki, R-P. \&
  Puls, J. 2000, \araa, 38, 613
\bibitem[Leavitt(1908)]{leavitt1908} Leavitt, H.S. 1908, Ann. Harvard
Coll. Obs. 60, 87
\bibitem[Marengo et al.(2001)]{marengo2001} Marengo, M., Karovska,
  M. Fazio, G. G., Hora, J. L., Hoffmann, W. F.,Dayal, A. \& Deutsch,
  L. K. 2001, \apj, 556, L47
\bibitem[Marengo et al.(2010)]{marengo2010} Marengo, M., Evans, N. R.,
  Barmby, P., Bono, G., Welch, D. \& Romaniello, M. 2010, \apj, 709,
  120
\bibitem[Mathis et al.(1977)]{mathis1977} Mathis, J. S., Rumpl, W. \&
  Nordsieck, K. H. 1977, \apj, 217, 425
\bibitem[Mathis et al.(1983)]{mathis1983} Mathis, J. S., Mezger,
  P. G. \& Panagia, N. 1983, \aap, 128, 212
\bibitem[McAlary \& Welch(1986)]{mcalary1986} McAlary, C. W. \& Welch,
  D. L. 1986, \aj, 91, 1209
\bibitem[M\'erand et al.(2006)]{merand2006} M\'erand, A. et al. 2006,
 \aap, 453, 155
\bibitem[M\'erand et al.(2007)]{merand2007} M\'erand, A. et al. 2007,
  \apj, 664, 1093
\bibitem[Moskalik et al.(1992)]{moskalik1992} Moskalik, P. Buchler,
  J.R. \& Marom, A. 1992, \apj, 385, 685
\bibitem[Natale et al.(2008)]{natale2008} Natale, G., Marconi, M.,
  Bono, G. 2008, \apj, 674, L93
\bibitem[Neilson \& Lester(2008)]{neilson2008} Neilson, H. R. \&
  Lester, J. B. 2008, \apj, 684, 569
\bibitem[Neilson et al.(2009)]{neilson2009} Neilson, H. R., Ngeow,
  C.-C., Kanbur, S. M. \& Lester, J. B. 2009, \apj, 692, 81
\bibitem[Ossenkopf \& Henning(1994)]{ossenkopf1994} Ossenkopf, V. \&
  Henning, T. 1994, \aap, 291, 943
\bibitem[Perryman et al.(1997)]{perryman1997} Perryman, M. A. C. et
  al. 1997, \aap, 323, 49
\bibitem[Raga \& Cant\'o(2008)]{raga2008} Raga, A. C. \& Cant\'o,
  J. 2008, \apj, 685, L141 
\bibitem[Rieke et al.(2004)]{rieke2004} Rieke, G. et al. 2004, \apjs,
  154, 25
\bibitem[Reimers et al.(1975)]{reimers1975} Reimers, D., 1975,
  Mem. Soc. Roy. Sci. Liege, 8, 369
\bibitem[Rogers \& Iglesias(1992)]{rogers1992} Rogers, F. J. \&
  Iglesias, C. A. 1992, \apj, 401, 316
\bibitem[Seaton et al.(1994)]{seaton1994} Seaton, M.J., Yan,
  Y. Mihalas, D. \& Pradhan, A. K. 1994, \mnras, 266, 805
\bibitem[Schuster et al.(2006)]{schuster2006} Schuster,
  M. T., Marengo, M. \& Patten, B. M. 2006, in Proc. SPIE 6270,
  Observatory Operations: Strategies, Processes, and Systems,
  ed. D. R. Silva \& R. E. Doxsey (Bellingham, WA: SPIE), 65 
\bibitem[Stobie(1969)]{stobie1969} Stobie, R. S. 1969, \mnras, 144,
  511
\bibitem[Storm et al.(2004)]{storm2004} Storm, J., Carney, B. W.,
  Gieren, W. P., Fouqu\'e, P., Latham, D. W. \& Fry, A. M. 2004, \aap,
  415, 531 
\bibitem[Ueta et al.(2006)]{ueta2006} Ueta, T. et al. 2006, \apjl, 648, L39
\bibitem[Ueta et al.(2008a)]{ueta2008a} Ueta, T. et al. 2008a, PASJ,
  60, 407
\bibitem[Ueta(2008b)]{ueta2008b} Ueta, T. 2008b, \apjl, 687, L33
\bibitem[van den Ancker(1997)]{ancker1997} van den Ancker, M. E.,
 Th\'e, P. S., Tjin A. Djie, H. R. E., Catala, C. de Winter, D.,
 Blondel, P. F. C. \& Waters, L. B. F. M. 1997, \aap, 324, L33
\bibitem[Vink et al.(2001)]{vink2001} Vink, J. S., de Koter, A. \&
  Lamers, H. J. G. L. M. 2001, \aap, 369, 574
\bibitem[Vitrichenko \& Tsarevskii(1969)]{vitrichenko1969}
  Vitrichenko, \'E. A. \& Tsarevskii, G. S. 1969, Soviet Astronomy,
  13, 159
\bibitem[Wareing et al.(2006)]{wareing2006} Wareing, C. J. et
  al. 2006, \mnras, 372, L63
\bibitem[Welch \& Duric(1988)]{welch1988} Welch, D. L. \& Duric,
  N. 1988, \aj, 95, 1794
\bibitem[Werner et al.(2004a)]{werner2004a} Werner, M. W. et al. 2004a,
  \apjs, 154, 1
\bibitem[Werner et al.(2004b)]{werner2004b} Werner, M. W., Uchida,
  K. I., Sellgren, K., Marengo, M., Gordon, K. D., Morriss, P. W.,
  Houck, J. R. \& Stansberry, J. A. 2004b, \apjs, 154, 309
\bibitem[Willson(2000)]{willson2000} Willson, L. A. 2000, \araa, 38,
  573 
\bibitem[Wilson(1953)]{wilson1953} Wilson, R. E. 1953, General Catalog
  of Stellar Radial Velocities, Washington (Carnegie Institute of
  Washington) 
\bibitem[Worley(1966)]{worley1966} Worley, C. E. 1966, \pasp, 78, 485 
\bibitem[Yong et al.(2000)]{yong2000} Yong, H., Demarque, P. \&
  Sukyoung, Y. 2000, \apj, 539, 928 





\end{thebibliography}
\end{document}